\documentclass[preprint,superscriptaddress,nofootinbib]{revtex4-2}
\usepackage{amsfonts}
\usepackage[T1]{fontenc}
\usepackage[latin9]{inputenc}
\usepackage{amsmath}
\usepackage{amssymb}
\usepackage{babel}
\usepackage{graphicx}
\usepackage{xcolor}%
\usepackage{braket}
 
\begin{document}
	\title{Trajectories in coupled waveguides: an application to a recent experiment and
		Hiley's lessons on the falsification of the Bohmian model }

	\author{F.\ Daem}
	\affiliation{Laboratoire de Physique Th\'eorique et Mod\'elisation, CNRS Unit\'e 8089,
	CY~Cergy~Paris~Universit\'e, 95302 Cergy-Pontoise cedex, France}

	\author{T. Durt}
	\affiliation{CLARTE, Aix Marseille Univ, CNRS, Centrale Marseille, Institut Fresnel, UMR 7249, Campus
		de St-J\'er\^ome, 13013 Marseille, France}

	\author{A. Matzkin}
	\affiliation{Laboratoire de Physique Th\'eorique et Mod\'elisation, CNRS Unit\'e 8089,
	CY~Cergy~Paris~Universit\'e, 95302 Cergy-Pontoise cedex, France}
	
	\begin{abstract}
		From \textquotedblleft surreal\textquotedblright\ trajectories to which-way
		measurements, Basil Hiley had a lesson: claims of falsifying the Bohmian model
		do not withstand scrutiny provided the model is applied correctly. In this
		work we compute de Broglie-Bohm trajectories for particles tunneling in
		coupled waveguides relevant to a recent experiment having claimed to challenge
		the Bohmian model. We show that the Bohmian model -- correctly applied --
		gives results identical to the standard quantum approach, first by working out
		a simple one-dimensional model, and then by computing Bohmian trajectories for
		the full two-dimensional problem representing a quantum particle propagating 
		inside coupled waveguides. We further
		recall the contextual nature of the Bohmian trajectories whereby the 
		trajectories of a closed system differ from the ones observed when an interaction with
		a measurement apparatus takes places.
		
	\end{abstract}
	\maketitle

\newpage

\section{Introduction}

The de Broglie-Bohm interpretation \cite{dB,bohm} has played an important role
in our understanding of quantum mechanics. Not that one should necessarily
endorse the ontological package and mechanisms put forward by the Bohmian
model as describing the \textquotedblleft real\textquotedblright\ behavior and
precise ontology of quantum systems.\
The strength of the Bohmian model lies, in our view, elsewhere: by proposing
an account of dynamical processes for which the orthodox interpretation tells
us we should give up any attempt to explain them, the Bohmian interpretation
improves our understanding of quantum phenomena, potentially leading to
important results, such as Bell's theorem \cite{hiley-bell}.

The issue of a possible falsification of the Bohmian approach seems to be a
settled issue, at least in the non-relativistic domain -- the standard
\emph{relativistic} quantum formalism is plagued with
interpretational problems (for example there is no agreement on the form of
the current density \cite{hoffmann} or on the existence of a position operator
\cite{daem-superradiance}) and unsurprisingly there is no generally accepted
formulation of a relativistic Bohmian model. Bohm and Hiley write, in the
introduction of their classic book "The Undivided Universe" \cite{undivided},
that "\emph{there is no way experimentally to decide}" between the standard
interpretation and the Bohmian model. Indeed, it is perfectly legitimate to
dislike the Bohmian account of a system's dynamics.\ For instance Einstein's
criticism of the Bohmian account of a particle in box ("That seems too
\emph{cheap} to me" \cite{Einstein-born}) is well-known. Such features might
make the Bohmian model less reasonable to some or might render the properties
of the trajectories less attractive in terms of explanatory power, but this
has no bearing on their empirical acceptability.

Nevertheless, from time to time, there are proposals that claim to falsify the
Bohmian approach.\ Such proposals might come in the form of ideal
thought-experiments, that would, if realized, render the Bohmian approach
untenable. Or perhaps ingeniously, there could be claims that an
\emph{experimentum crucis} falsifying the Bohmian model was realized. Basil
Hiley would deconstruct such claims showing where the Bohmian model was
incorrectly applied. This would be done patiently, calmly working out each
step, in the same clear way he would discuss in person with anyone\footnote{One 
	of us (A.M.) wishes to recall in this memorial volume the first
and last times he met Basil Hiley. ``\emph{My first encounter
with Basil was... on the day of his retirement (in 2001)! I was working in
London back then, and I frequently had lunch with Peter Van Reeth, a fellow
UCL physicist. We would embark during lunch on discussions ranging from the
foundations of quantum mechanics to the sociology of science.\ Peter was
informed that a small meeting was organized at Birkbeck on the occasion of
Basil's retirement.\ We walked the couple of blocks separating UCL from
Birkbeck, and slipped into the room where the event was organized (and in some
sense, into an entirely different world). Years later (2023), I was
co-organizer (in particular with another co-author of this paper, T.D.) of a
conference commemorating 100 years of Louis de Broglie's first publications on
quantum mechanics. Basil was of course invited. He was slightly impaired
physically, but was extremely sharp intellectually.\ He delivered an impressive
and energetic lecture.}''}. It is worth recalling here his
efforts in debunking the surreal trajectories proposal \cite{surreal,scully}. The
main idea behind surreal trajectories relies on the no-crossing rule by which
Bohmian trajectories abide\footnote{The simplest setup to see this is the
following one-dimensional case \cite{1DBW}: imagine an initial state as a superposition of
two identical Gaussian wavepackets, one sitting by Alice's side, the other by
Bob's.\ Alice launches her wavepacket towards Bob, and Bob sends his
wavepacket to Alice.\ While the two wavepackets cross (Alice's wavepacket
reaching Bob and vice-versa), the Bohmian trajectories turn back in the
crossing region, so that Alice detects the same Bohmian particle that was
initially launched by her (and similarly for Bob if his detector is the
one that fires).}. Hiley discussed this problem and the associated
\textquotedblleft which way\textquotedblright\ problem in details \cite{hm,eraser,welcher,peter-surreal}
and would typically conclude, as in the Conclusion of
Ref.\ \cite{peter-surreal}: \textquotedblleft\emph{we have shown that the
differences that are claimed to exist between the standard approach to quantum
mechanics and the Bohm approach do not exist when both are applied correctly.
Indeed it is hard to imagine how there could be any differences in the
predicted experimental results since both approaches use exactly the same
mathematical structure}\textquotedblright.

The same conclusion should apply to a recently published experimental work
\cite{exp} that claims to have falsified the Bohmian model by comparing
observed particle populations tunneling through coupled waveguides to what the
authors of that work expect Bohmian particles would do, namely remain
stationary in the tunneling region. In this work, we show below that this
expectation, that will appear as obviously problematic to anyone that has some
experience with Bohmian dynamics, is indeed incorrect.\ First (Sec.
\ref{sec:1D}) we will introduce a simple one dimensional (1D) double-well
model, relevant to the experiment reported in \cite{exp} to show that generic
Bohmian trajectories in that region are not stationary. Then in Sec.
\ref{sec:2D} we solve numerically the 2D Schr\"odinger equation corresponding to
the coupled waveguides of the experiment, and compute the corresponding
Bohmian trajectories. We then briefly discuss in Sec. \ref{sec:DC} the
particular case of stationary states, and conclude
on the impossible falsification of the Bohmian model -- which is at the same time a
strength (in that standard results are always recovered) and a weakness (in
that there can be no observational warrant to give a credible empirical
status to its ontological package).

\section{De Broglie-Bohm trajectories for a particle propagating between two
coupled potential wells}
\label{sec:1D}

\subsection{Experimental observation of tunneling between coupled waveguides}

\begin{figure}[h]
	\centering
	\includegraphics[width=1\textwidth]{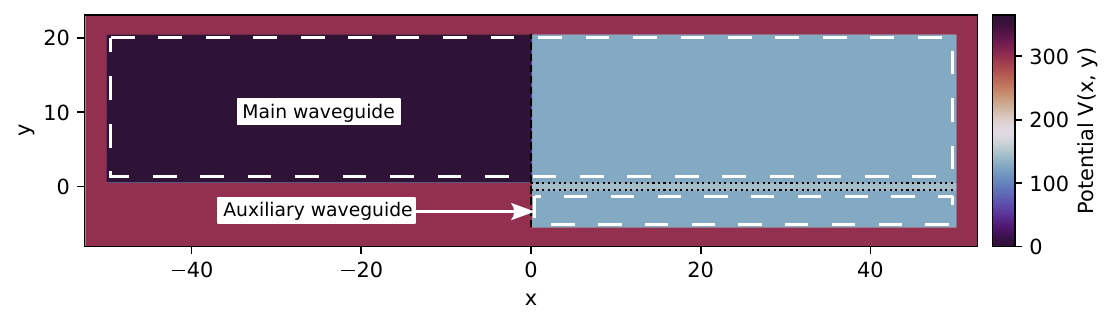}
	\caption{Main and auxiliary waveguides displayed with a height map representing the potential $V(x,y)$
		(the height scale corresponds to the numerical computations of Sec. \ref{sec:2D}).
		The step potential  (in the region $x > 0$) and the barrier potential along $y$ between the two
		waveguides are clearly visible.}
	\label{fig:potential_2d}
\end{figure}

A recent optical microcavity experiment \cite{exp} measured the photon
population transfer between two coupled waveguides (a main one and an
auxiliary one). Photons are fed into the main waveguide and encounter a step
potential before being coupled to the auxiliary waveguide by way of an
additional barrier potential (see Fig. \ref{fig:potential_2d}). It is claimed in
Ref. \cite{exp} that these observations contradict the Bohmian model because,
according to the authors, the Bohmian particles should not move once they are
inside the potential step, and could therefore not account for the transfer
between the two waveguides.

We will not be interested here in the specific experiment carried out in Ref.
\cite{exp}.\ Indeed, that experiment involves photons, and even if an
effective two-dimensional Schr\"odinger-like equation can be derived, in the
paraxial approximation, for photons in a planar waveguide fundamentally
photons are relativistic bosons and there is no consensus
on whether they can be described as following Bohmian
	trajectories (see eg Ch.\ 11 of \cite{undivided} or \cite{thomFP1}). We will instead take the same
setup and assume a massive quantum particle obeys the Schr\"odinger equation in the same
2D potential $V(x,y)$ scheme of the experiment \cite{exp}, and determine the corresponding Bohmian trajectories, that will not turn out
to correspond to static particles.

The potential $V(x,y)$ is mainly characterized by a step potential along
$x$ (of height $V_{step}$ for $x>0$) and, in the region $x>0$, a barrier potential along $y$ of height $V_{barrier}$ and width $d$  (centered at
$y=0$) separating the two waveguides (see Fig. \ref{fig:potential_2d}). The combination of these two potentials
for the waveguides can be written ($\theta$ is the step function) as
\begin{equation}
V_{in}(x,y)= \theta(x)\left(V_{step}
+\theta(d/2-|y|)V_{barrier} \right) \, ,
\end{equation}
and the total potential includes in addition the boundaries of the waveguides represented by $V_{out}(x,y)$ so that

\begin{equation}
	V(x,y)=V_{in}(x,y)+ V_{out}(x,y)\,.
	\label{pot-def}
\end{equation}

Before doing the necessarily numerical
computations for particle trajectories in this potential, we examine a much simpler 1D model that lies
at the heart of the population transfer observed experimentally.

\subsection{Bohmian trajectories tunneling between two wells: a 1D model}

It is well-known that when the Hamiltonian is time independent and that the
wavefunction of the system is real, the de Broglie-Bohm (dBB)
velocity is equal to 0. This is so for instance for discrete eigenstates of
the Hamiltonian e.g. the electronic ground state of the hydrogen atom, or for
the energy eigenstates of a particle in a box. In tunneling
situations, when the potential is constant in a region of space (the potential
barrier) and the energy of the quantum system is smaller than the potential
the eigenstates of the Hamiltonian inside the potential are locally real
exponential functions (if the potential range is semi-bounded) 
or a (possibly complex) combination of real exponential functions,
and for those states the dBB velocities are predicted to
be equal to 0 (in the former case) or constant.

Our aim here is to show that for the potential along $y$ (see Fig. \ref{fig:potential_2d})
that couples both waveguides the dBB velocities
are in general \emph{not} equal to zero in the tunnel zone. To illustrate this
point we analyse a problem that is treated in several textbooks, tunneling
between two 1D square-well potentials. To do so, we consider a particle of
mass $m$ trapped in a 1D potential equal to $-V_{0}$ (with $V_{0}$ a positive
real number) in the regions $-d/2-a\leq y\leq-d/2$ and $d/2\leq y\leq d/2+a$,
defining the two wells of width $a$; the potential is 0 elsewhere. To prove that dBB
velocities are not equal to 0 in the tunneling zone it is sufficient as we
shall show now to consider a coherent superposition of two energy level
states. Moreover we show explicitly that at all times the statistical
distribution of positions obeys the Born rule for such a system, an
illustration that it is not possible to falsify the dBB interpretation.

\subsubsection{Lowest energy levels.}

\begin{figure}[h]
\centering
\includegraphics[scale=0.45]{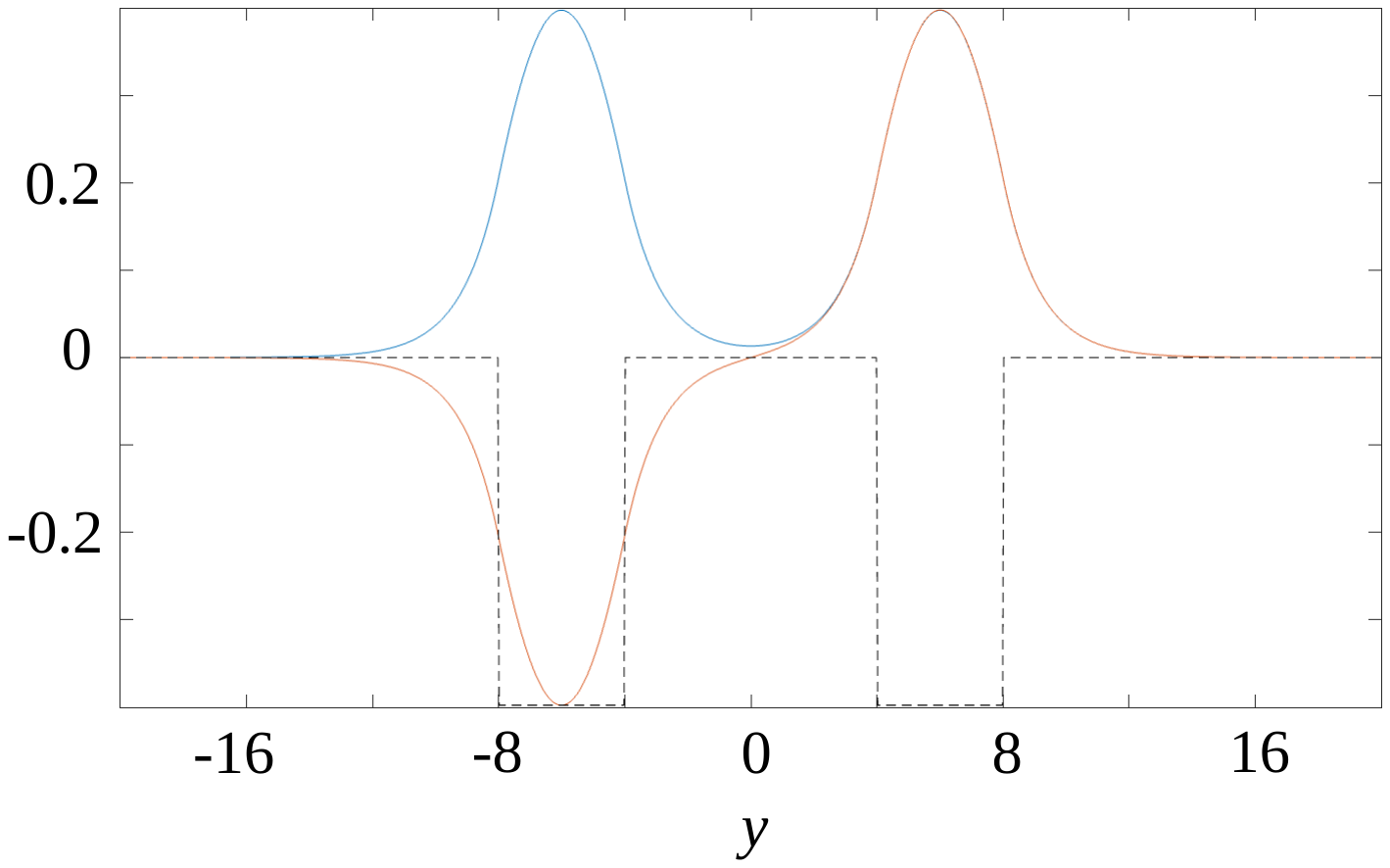}
\caption{Schematics showing the two lowest energy levels in a
double well represented with the dashed lines: fundamental level in blue, first excited level in orange, with a
sinusoidal behaviour inside the wells, and an exponential behaviour outside (atomic units are used; the left scale refers to the wavefunction amplitudes).}%
\label{fig2}%
\end{figure}

As the potential is symmetric around the origin of the $y$ axis, one can
predict that the lowest energy level (of energy $E_{0}$) is associated in the
position representation with an even function of the position and the first
excited level (of energy $E_{1}$) with an odd function (Fig. \ref{fig2}).
These levels are determined, as is usually done for simple 1D scattering
problems, by looking for superpositions of real exponential functions in the
classically forbidden regions (where the energy is smaller than the potential)
and superpositions of plane waves in the classically allowed regions.


Imposing continuity of the wave function and of its first derivative and also
imposing that when $|y|$ goes to infinity the wave function is a decreasing
exponential function leads to a
quantization of the energy levels. The quantization conditions are expressed
through transcendental equations that can be solved numerically. Here we
choose for convenience deep potential wells separated by a \textquotedblleft
large\textquotedblright\ (see below) distance, so that in each well the
influence of the other well is small and the states should be close to the
solutions inside an infinite well, $\sqrt{{\frac{1}{a}}}sin({\frac{\pi}{a}%
}(y\pm d/2))$ for the left and right wells resp.

We therefore look for solutions for the ground state and first excited level
having inside the wells the form $\ket{\psi_{0}}\approx N\cdot sin({\frac{\pi}{a}%
}(y+d/2))$ in the left well and $\ket{\psi_{0}}\approx N\cdot sin({\frac{\pi}{a}%
}(y-d/2))$ in the right well ($N$ being a normalisation factor close to
${\frac{1}{\sqrt{a}}}$), while $\ket{\psi_{1}}\approx N\cdot sin({\frac{\pi}{a}%
}(y+d/2))$ in the left well and $\ket{\psi_{1}}\approx-N\cdot sin({\frac{\pi}{a}%
}(y-d/2))$ in the right well. The energies $E_{0}$ and $E_{1}$ are close to
each other and also close to the energy of the fundamental state inside a
unique very deep well. In the barrier region the fundamental and first excited
states are respectively equal (up to a normalisation factor to be specified
later) to $ch(y/L_{0})$ and $sh(y/L_{1})$ where $-{\hbar^{2}/}2mL_{i}%
^{2}=E_{i}$, for $i=0,1$.
These typical behaviours are visible in Fig. \ref{fig2}. 

\subsubsection{Tunnel oscillations}

Let us now focus on the region between the two wells where the trapped
particle exhibits tunnel oscillations. Let us define the \textquotedblleft
left\textquotedblright\ and \textquotedblleft right\textquotedblright\ wave
functions as follows:%

\begin{equation}
\ket{\psi_{L,R}}={\frac{1}{\sqrt{2}}}(\ket{\psi_{0}} \pm\ket{\psi_{1}}).
\end{equation}

In the case of deep wells ( ${\frac{\hbar^{2}\pi^{2}}{2ma^{2}}\ll}V_{0}$)
separated by a large distance ($L_{i}{\ll d}$) the support of the left (right)
wavefunction is essentially confined inside the left (right) well, with a real
exponential function in between the wells rapidly decreasing at the right
(left) of the left (right) well and an exponential decay at infinities. The
weight of these real exponential functions is very small because they rapidly
decrease and also because the continuity conditions impose that their weights
at the edge of the wells are very small.

Let us assume that at time $t=0$ we confine the particle in the left well by
preparing the state $\ket{\psi(t=0)} =\ket{\psi_{L}} .$ Then in our 2 level model, $\ket{\psi(t)}=e^{-i{\frac{E_{0}+E_{1}}{2\hbar}}t}(e^{-i{\frac{E_{0}-E_{1}}{2\hbar}}t}\ket{\psi_{0}}+\ket{e^{i{\frac{E_{0}-E_{1}}{2\hbar}}t}\ket{\psi_{1}}})/\sqrt{2}.$
Defining $\omega_{tunnel}={\frac{E_{0}-E_{1}}{2\hbar}}$, and making use of the
relations $\ket{\psi_{0}}={\frac{1}{\sqrt{2}}}(\ket{\psi_{L}}+\ket{\psi_{R}})$ and
$\ket{\psi_{1}}={\frac{1}{\sqrt{2}}}(\ket{\psi_{L}}-\ket{\psi_{R}})$, we find that%

\begin{equation}
\left\vert \psi(t)\right\rangle =e^{-i{\frac{E_{0}+E_{1}}{2\hbar}}t}%
(\cos(\omega_{tunnel}t)\ket{\psi_{L}}-i\cdot\sin(\omega_{tunnel}t)\ket{\psi_{R}}).
\end{equation}
In other words, the system exhibits tunnel oscillations between the two wells.
As we consider here that the wells are deep and separated by a large distance,
these oscillations are slow because the difference between the energies of the
two lowest energy states is small.

\subsubsection{Tunneling current density and non-zero Bohmian velocities}

The state $\ket{\psi(t)}$ fulfills Schr\"odinger's equation
in between the two wells so that the Born density $|\psi(y,t)|^{2}$ obeys the
conservation equation ${\frac{\partial}{\partial t}}|\psi(y,t)|^{2}$%
+${\frac{\partial}{\partial y}}J_{y}(y,t)=0$, where
\begin{equation}
J(y,t)={\frac{\hbar}{m}}\operatorname{Im}\left[  \psi^{\ast}(y,t)\partial
_{y}\psi(y,t)\right]  . \label{1d-dens}%
\end{equation}
Starting from $|\psi(y,t)|^{2}=|\psi(0,0)|^{2}\cdot|ch{\frac{y}{L_{tunnel}%
^{0}}}e^{-i{\frac{E_{0}}{\hbar}}t}+sh{\frac{y}{L_{tunnel}^{1}}}e^{-i{\frac
{E_{1}}{\hbar}}t}|^{2}$, and having in mind that we are dealing with deep and
distant wells so that $E_{0}$ is very close to $E_{1}$ and hence
($E_{i}=-{\hbar^{2}/}2mL_{i}^{2}$) $L_{tunnel}^{0}\sim L_{tunnel}^{1}\sim
L_{tunnel}$ , we obtain%
\begin{equation}
{\partial}_{y}J(y,t)=(-4)|\psi(0,0)|^{2}\omega_{tunnel}\sin(2\omega
_{tunnel}t)ch\left(  {\frac{y}{L_{tunnel}^{0}}}\right)  sh\left(  {\frac
{y}{L_{tunnel}^{1}}}\right)  . \label{j-1d}%
\end{equation}

Now recall that by definition the local de Broglie-Bohm velocity is given by
\begin{equation}
v(y,t)=\frac{J(y,t)}{|\psi(y,t)|^{2}}%
\end{equation}
and therefore per Eq. (\ref{j-1d}) does not vanish in the tunnel region in
between the two wells, which is the main result of this section. Actually if
we put
$L_{tunnel}^{0}=L_{tunnel}^{1}=L_{tunnel}$
we get
\begin{equation}
	J_{y}\approx
|\psi(0,0)|^{2}{\frac{\hbar}{m}}\sin({\frac{(E_{1}-E_{0})}{\hbar}%
}t)/2L_{tunnel}\,,
\end{equation}
which indicates that the current takes the same value, at a
given time $t$, everywhere inside the tunnel zone.

It is worth emphasizing that the reasoning made above can be generalized to
any coherent superposition of energy levels. The distribution of positions is
directly linked to the wavefunction and thus obeys the Born rule everywhere in
space, and at all times, which renders the dBB dynamics indistinguishable from
the one resulting from the Schr\"odinger equation. This holds irrespective of
whether the underlying basis is composed of oscillating or evanescent waves,
and has been known for a long time in the context of tunneling dynamics
\cite{muga}: the standard and dBB results are identical despite the
wavefunction being composed of evanescent waves in the barrrier region.

We also note that at the edges of the tunneling region the current as
 well as the dBB velocity are continuous
functions of $y$. Hence since the variation of
the probability to be in, say, the left well varies
as ${\frac{-d}{dt}%
}cos^{2}(\omega_{tunnel}t)$ and is equal to the current density passing
from the well to the tunnel zone, the Bohmian particles cannot remain static.

\section{Computation of 2D Bohmian trajectories in coupled waveguides}
\label{sec:2D}

We now turn to the 2D problem with the potential given by Eq.
(\ref{pot-def}) and compute the Bohmian trajectories for a wavepacket
propagating in the main waveguide and subsequently tunneling to the auxiliary
waveguide. 

We represent the borders of the waveguides by the potential $V_{out}(x,y)$  defined in terms of step functions such that $V_{out}$ is equal to a very high $V_o$ outside the waveguides and zero inside the regions of the waveguides ($d/2<y<d/2+a$ for $-L/2<x<L/2$ and $-d/2-b<y<-d/2$ for $0<x<L/2$, where $a$ and $b$ are the respective width of the main and auxiliary waveguides, and L the length of the first waveguide).

The particle is represented by a Gaussian wavepacket of initial momentum $p_{0}$ and spatial width
$\sigma$ centered around the position $x_{0}$. It is expressed as
\begin{equation}
	\psi(x,y,t=0)=N e^{-\frac{(x-x_{0})^{2}}%
		{2\sigma^{2}}}e^{i\frac{p_{0}(x-x_{0})}{\hbar}}\phi(y),
\end{equation}
where $N$ is a normalisation factor and $\phi(y)$ could be for instance a superposition
of the ground and first excited states of the one dimensional double well (see Sec. \ref{sec:1D}).
For simplicity we will take $\phi(y)$ as a Gaussian of same width centered around the main
waveguide and with zero initial momentum along $y$:
\begin{equation}
	\phi(y)=e^{-\frac{(y-y_{0})^{2}}{2\sigma^{2}}},
\end{equation}
where $y_{0}$ is the center of the main waveguide.

The numerical simulation region contains the two waveguides (see Fig. \ref{fig:potential_2d}). Atomic units are used throughout the simulation ($\hbar=m=1$).
The main waveguide is
of total length $L=100\,\mathrm{a.u.}$ and width $a=20\,\mathrm{a.u.}$, and the auxiliary waveguide is of 
length $L/2$ but of width $b=5\,\mathrm{a.u.}$ only. A smaller width is taken for the auxiliary
waveguide to reduce the computational cost of the simulation while still
allowing for tunneling effects to be captured. For numerical purposes we take a smoothed version of the step potentials
by replacing all  step functions $\theta(x)$ by hyperbolic tangents
of the form $(1+\tanh(x/\epsilon))/2$ with $\epsilon$ small enough (we took
$\epsilon=0.05\,\mathrm{a.u.}$). For the potentials, we have taken $V_o=10^4\,\mathrm{a.u.}$, $V_{step}=162\,\mathrm{a.u.}$, $V_{barrier}=18\,\mathrm{a.u.}$ with
 $d=1\,\mathrm{a.u.}$.

The wavepacket is then propagated numerically by solving the 2D time-dependent Schr\"odinger
equation with the potential of Eq. (\ref{pot-def}) using a split-operator
algorithm. The particle is evolved up to a final time $t_{f}=5\,\mathrm{a.u.}$ with
a time step $\delta t=10^{-4}\,\mathrm{a.u.}$. The discretization steps in space are $\delta x \approx\delta y \approx 0.03\,\mathrm{a.u.}$ (corresponding to 3072 $x$ points and 1024 $y$ points).

\begin{figure}
	\centering
	\includegraphics[width=0.8\textwidth]{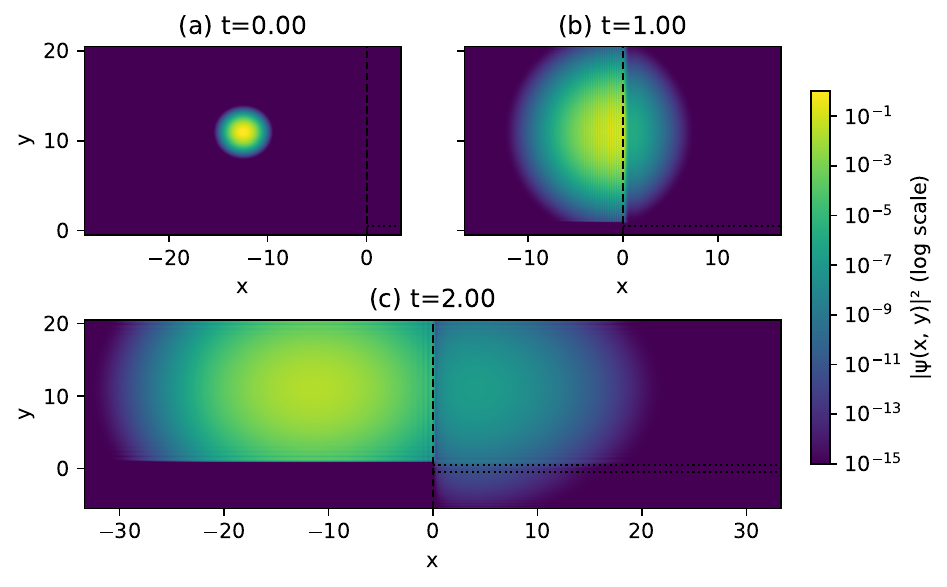}
	\caption{Time evolution of the probability density $|\psi(x,y,t)|^2$.
		At time $t=0\,\mathrm{a.u.}$ (a) the wavepacket is centered around $x_0=-12.5\,\mathrm{a.u.}$ and $y_0=10.5\,\mathrm{a.u.}$, with width $\sigma=0.5\,\mathrm{a.u.}$, momentum $p_0=12\,\mathrm{a.u.}$ along $x$ and
		no initial momentum along $y$.
		At time $t=1\,\mathrm{a.u.}$ (b) the wavepacket is reflected and transmitted at the step potential (at $x=0\,\mathrm{a.u.}$).
		At time $t=2\,\mathrm{a.u.}$ (c) the transmitted part of the wavepacket starts to tunnel to the auxiliary waveguide.
		We use a logarithmic color scale to enhance the visibility of the small
		transmitted density.
		The step potential (at $x = 0\,\mathrm{a.u.}$) is indicated by the dashed line, and the barrier potential (between the two waveguides) is indicated by the dotted lines.
		Atomic units ($\mathrm{a.u.}$) are used throughout ($\hbar=m=1$).}
	\label{fig:evolution}
\end{figure}

The wavepacket is initially
located in the main waveguide (Fig. \ref{fig:evolution} (a)), away from the step potential, and propagates
towards the step potential. The wavepacket is reflected and transmitted at
the step potential (Fig. \ref{fig:evolution} (b)), and the transmitted part subsequently tunnels to the
auxiliary waveguide (Fig. \ref{fig:evolution} (c)). In this two dimensional model, the density tunneling
towards the auxiliary waveguide is understood to come from the spreading
of the wavepacket along $y$ as it propagates along $x$. The current density
is clearly non-zero in the tunneling region, and we expect the Bohmian
particles to move accordingly.

The Bohmian trajectories are computed by integrating the velocity field
\begin{equation}
\mathbf{v}(x,y,t)=\frac{\hbar}{m}\operatorname{Im}\left[  \frac{\nabla
\psi(x,y,t)}{\psi(x,y,t)}\right]
\end{equation}
at each time step of the simulation.

\begin{figure}[h]
	\centering
	\includegraphics[width=0.8\textwidth]{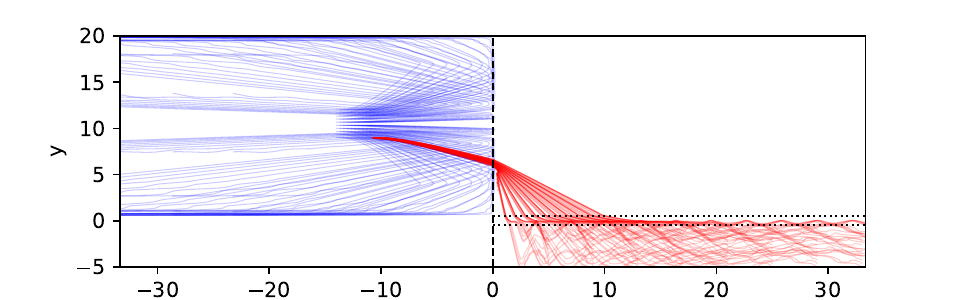}
	\caption{Bohmian trajectories computed for the 2D potential of Eq. (\ref{pot-def}).
		Two sets of trajectories are shown: the first set (in blue) corresponds to particles that are initially located around the initial probability density, and that reflect off the step potential.
		The second set (in red) corresponds to particles that are taken to be regularly spaced in the auxiliary waveguide at $t=t_f=5\,\mathrm{a.u.}$.
		The step potential (at $x = 0\,\mathrm{a.u.}$) is indicated by the dashed line, and the barrier potential (between the two waveguides) is indicated by the dotted lines.
		Atomic units ($\mathrm{a.u.}$) are used throughout ($\hbar=m=1$).}
	\label{fig:bohmian_trajectories}
\end{figure}

We present in Fig. \ref{fig:bohmian_trajectories} two sets of typical Bohmian trajectories obtained from this simulation. The first set of (more probable) trajectories corresponds to particles that are initially located around the initial probability density, and that reflect off the step potential. This represents the
main part of the wavepacket that is reflected at the step potential.
The second set corresponds to (less probable) particles that end up in the auxiliary waveguide after tunneling. The Bohmian particles are clearly not static in the tunneling region, and the trajectories account for the population transfer to the auxiliary waveguide.

\section{Discussion and Conclusion}
\label{sec:DC}

We have first looked at a 1D model (accounting for coupling between two wells)
and saw that for the simplest 2-level state the Bohmian particles are not
stationary. We then computed dBB trajectories for the 2D case describing the
propagation of a wavepacket in the main guide and the subsequent coupling to the auxiliary
guide as observed experimentally. The Bohmian particles are not static and account for populating the
auxiliary cavity. These results illustrate that the Bohmian analysis of the
experiment reported in Ref. \cite{exp} is incorrect.\ Any sensible modeling of
an experiment will involve wavepackets and non-zero current densities and
Bohmian velocities.

Nevertheless, what happens for genuine stationary states? Generically Bohmian
particles have a constant velocity, which can be zero for bounded or
semi-bounded systems, like the step potential applied here to the waveguides in the region $x>0$. This is not surprising, and
reflects the fact that the Schr\"odinger current density is constant or
vanishing in such cases. Hence there is no mismatch between the dBB and
standard approaches, though the issue of static Bohmian particles has often been discussed in
interpretational terms. 

In the dBB approach it is crucial to remark that the static Bohmian particles for real stationary states are relevant to a 
\textit{closed system}; when the system is measured, the interaction Hamiltonian always modifies the
closed system trajectories, so that the static Bohmian particles cannot be
observed in principle. This contextual character of Bohmian 
trajectories is well-known and has been discussed at great lengths for the particle in a box (see
Ch.\ 6 of Bohm and Hiley's book \cite{undivided}, or Sec.\ 6.5 of
ef.\ \cite{holland} where the effect of a measurement on the closed-system static particle is analyzed, so
that for instance a measurement of the particle momentum is never zero).
This feature has also been discussed for more involved systems, such as quantum billiards
\cite{billiard}, in which the eigenstates (and the observed time-dependent dynamics) are organized along classical
trajectories though the closed-system dBB trajectories are unrelated to these
classical trajectories. This mismatch between an unobservable closed system behavior and
the Bohmian trajectories that appear when a measurement takes place, analyzed in \cite{mismatch}, 
could be taken as a valid epistemological criticism of the dBB interpretation, but it has
no implication on the falsification of the model, as the standard quantum statistics are always recovered.

The upshot concerning the Bohmian analysis proposed by the authors of Ref.
\cite{exp} is that it is inconsistent to analyze the Bohmian model in terms
of a hand-waving semiclassical approach based on a single stationary plane wave for a
closed, non-observed system, while describing an experimental observation that
hinges on a rich time-dependent quantum dynamics.\ Perhaps this would have
been the geist of Basil Hiley's remarks. Note that
the	experiment \cite{exp} was described from a Bohmian viewpoint in two brief 
comments. Computations for a 1D two-level system given in Ref. \cite{exp}
but somewhat perplexingly not employed to discuss the Bohmian approach in that paper were
given in Ref. \cite{r1}, showing, as our model given in Sec. \ref{sec:1D}, non-static
Bohmian particles, while a model based on a 2D ansatz for the
wavefunction with a 1D effective potential (along $y$) reaches
the same conclusions \cite{r2} and displays 2D Bohmian trajectories qualitatively similar
to those we obtained in Sec. \ref{sec:2D}.

Another conclusion is that the fact that the de Broglie-Bohm model cannot be
falsified is a strength of the interpretation, as it can always offer a
consistent account to understand the underlying dynamics explaining the
behavior of a quantum system and the corresponding observations. But it is
also a weakness to the extent that the Bohmian model offers a candidate
ontology that is bound to remain an eternal candidate: there are no
observational warrants able to distinguish the model from standard quantum
mechanics that could indicate, albeit in an indirect way, that particles and
pilot waves are out there, in the physical world. Perhaps however
the situation becomes
more interesting if we follow Basil Hiley into believing that at best, the de
Broglie-Bohm model is a simplified and partial view of some deeper reality.
Then one might come up with models \cite{thomFP2} encompassing the pilot-wave in some limit,
but leading to different and -- in this case -- falsifiable predictions.


\begin{thebibliography}{99}                                                                                               %
\bibitem {dB}L. De Broglie, J. Phys. Radium, 8, 225 (1927).

\bibitem {bohm}D. Bohm, Phys. Rev. 85, 166 (1952) ; Phys. Rev. 85, 180 (1952).

\bibitem {hiley-bell}B.\ J.\ Hiley, Some Personal Reflections on Quantum
Non-locality and the Contributions of John Bell, arXiv:1412.0594v1
[physics.hist-ph], 2014.

\bibitem {hoffmann}S. E Hoffmann, J. Phys. A: Math. Theor. 52 225301 (2019).

\bibitem {daem-superradiance}F. Daem and A. Matzkin, Effects of superradiance on relativistic Foldy-Wouthuysen densities Phys. Rev. A 111, L060202, 2025

\bibitem {undivided}D.\ Bohm and B.\ J.\ Hiley, \textit{The Undivided
	Universe} (Routledge, London, 1993).

\bibitem {Einstein-born}A. Einstein, in M. Born and A. Einstein, The Born-Einstein letters (Macmillan, 1971), p. 192.
\bibitem {surreal}Englert, J.; Scully, M.O.; S�ssman, G.; Walther, H.
Surrealistic Bohm Trajectories. Z. Naturforsch. 47, 1175 (1992).

\bibitem {scully}Scully, M. Do Bohm trajectories always provide a trustworthy
physical picture of particle motion? Phys. Scr. , 76, 41--46 (1998).

\bibitem {1DBW} F. Daem and A. Matzkin, Dynamics, locality and weak measurements: trajectories and which-way information in the case of a simplified double-slit setup
 Quantum Stud.: Math. Found. 11, 567, 2024

\bibitem {hm}Hiley, B. J., Callaghan, R. E. and Maroney, O., Quantum
Trajectories, Real, Surreal, or an Approximation to a Deeper Process.
quant-ph/0010020, (2000).

\bibitem {eraser}B. J. Hiley and R. E. Callaghan, What is erased in the
quantum erasure? Found.\ Phys.\ 36, 1869 (2006).

\bibitem {welcher}B.\ J.\ Hiley, Welcher Weg Experiments from the Bohm
Perspective, AIP Conf. Proc. 810, 154 (2006).

\bibitem {peter-surreal}B. J Hiley and P. Van Reeth Quantum
Trajectories: Real or Surreal? Entropy, 20, 353 (2018).

\bibitem {exp}Sharoglazova, V., Puplauskis, M., Mattschas, C., Toebes, C. \&
Klaers, J. Energy--speed relationship of quantum particles challenges Bohmian
mechanics. Nature 643, 67 (2025).

\bibitem {thomFP1} T. Durt, Do (es the influence of) empty waves survive
 in configuration space? Found. Phys. 53, 13, 2023.


\bibitem {muga}J.G. Muga and C.R. Leavens, Arrival time in quantum mechanics,
Phys.\ Rep.\ 338, 353 (2000).


\bibitem {holland}P. R. Holland, \emph{The Quantum Theory of Motion}
(Cambridge University Press, Cambridge, England, 1993).

\bibitem{billiard}A. Matzkin, Bohmian mechanics, the quantum-classical correspondence and the classical limit: the case of the square billiard, Found. Phys. 39, 903, 2009. 

\bibitem{mismatch}A. Matzkin and V. Nurock, Classical and Bohmian trajectories in semiclassical systems: Mismatch in dynamics, mismatch in reality?, Stud. Hist. Phil. Mod. Phys. 39, 17, 2008. 

\bibitem{r1}Y.-F. Wang, X.-Y. Wang, H. Wang and C.-Y. Lu, Tunnelling photons pose no challenge to Bohmian machanics,  	arXiv:2507.20101 [quant-ph], 2025.

\bibitem{r2}A. Drezet, D. Lazarovici and B. M. Nabet, Comment on "Energy-speed relationship of quantum particles challenges Bohmian mechanics",  	arXiv:2508.04756 [quant-ph], 2025.

 
\bibitem {thomFP2}T. Durt, Testing de Broglie's Double Solution in the Mesoscopic Regime, Found. Phys. 53, 2, 2023.


\end{thebibliography}
\end{document}